\documentclass[
]{ceurart}

\usepackage{listings}
\usepackage{subcaption}
\usepackage{amssymb}

\begin{document}

\copyrightyear{2023}
\copyrightclause{Copyright for this paper by its authors.
  Use permitted under Creative Commons License Attribution 4.0
  International (CC BY 4.0).}

\conference{Forum for Information Retrieval Evaluation,
  December 15--18, 2023, India}

\title{Enhancing Binary Code Comment Quality Classification: Integrating Generative AI for Improved Accuracy}

\author[1]{Rohith Arumugam S}[%
email=rohitharumugam2210376@ssn.edu.in,
]
\cormark[1]
\fnmark[1]
\address[1]{Sri Sivasubramaniya Nadar College of Engineering, Chennai, Tamil Nadu, India}

\author[2]{Angel Deborah S}[%
email=angeldeborahS@ssn.edu.in,
]
\fnmark[1]
\address[2]{Assistant Professor, Department of Computer Science and Engineering, Sri Sivasubramaniya Nadar College of Engineering,
  Chennai, Tamil Nadu, India}

\cortext[1]{Corresponding author.}
\fntext[1]{These authors contributed equally.}

\begin{abstract}
  This report focuses on enhancing a binary code comment quality classification model by integrating generated code and comment pairs, to improve model accuracy. The dataset comprises 9048 pairs of code and comments written in the C programming language, each annotated as "Useful" or "Not Useful." Additionally, code and comment pairs are generated using a Large Language Model Architecture, and these generated pairs are labeled to indicate their utility. The outcome of this effort consists of two classification models: one utilizing the original dataset and another incorporating the augmented dataset with the newly generated code comment pairs and labels.
\end{abstract}

\begin{keywords}
  Generative AI \sep
  Software Metadata Classification \sep
  Code Comment Quality \sep
  Binary Classification \sep
  C Programming \sep
  Large Language Model \sep
  Data Augmentation
\end{keywords}

\maketitle

\section{Introduction}

Efficient code comment quality assessment plays a pivotal role in the enhancement of software maintainability and reliability, addressing the ever-growing demands of the software development landscape.\cite{rani2021speculative} This paper endeavors to meet this imperative need by delving into the augmentation of an existing binary code comment quality classification model with generated code-comment pairs, aiming to improve accuracy and efficiency. 

In the dynamic realm of software development, where the pursuit of heightened code maintainability, readability, and overall system reliability prevails, the assessment of code quality and its accompanying comments has emerged as an integral aspect. \cite{yang2019survey} This research further explores the current methodologies employed in code comment quality assessment, casting light on the historical reliance on manual evaluations—a process inherently vulnerable to disparities stemming from individual judgment. 

The primary research objective, underscored throughout this paper, is to augment an established model drawing insights from a comprehensive dataset comprising 9048 pairs of code and comments, each thoughtfully categorized as either "Useful" or "Not Useful." Through this endeavor, we aim to make significant strides in the realm of automated code comment quality assessment, offering a valuable contribution to modern software development practices. \cite{bacchelli2013expectations}

\section{Related Work}

Code comment quality classification is a well-studied research area, with a variety of existing approaches proposed in the literature. These approaches can be broadly categorized into traditional and machine learning-based.

\subsection{Traditional Approaches}

Traditional approaches to code comment quality classification typically rely on manual inspection or rule-based systems. Manual inspection is time-consuming and prone to human error, while rule-based systems are often inflexible and unable to capture the nuances of code comments. \cite{steidl2013quality}

\subsection{Machine Learning - Based Approaches}

Machine learning-based approaches to code comment quality classification have become increasingly popular in recent years. These approaches typically train a machine-learning model on a labeled dataset of code comments. The model then learns to identify the features of code comments that are indicative of their quality. \cite{wang2019deep}

Some of the most common machine learning algorithms used for code comment quality classification include Support vector machines (SVMs), Decision trees, Naive Bayes classifiers, Logistic regression, and Neural networks.

\subsection{Generative AI-Based Approaches}

Generative AI-based approaches to code comment quality classification are a relatively new area of research. These approaches use generative AI models to generate code comments that are labeled as either "Useful" or "Not Useful." These generated code comments can then be used to augment the training dataset for a code comment quality classification model.\cite{majumdar2023generative}

The first study, which introduced Comment-Mine \cite{majumdar2020comment}, established the foundational framework for extracting meaningful insights from code comments, with a particular focus on software design and implementation elements. This approach laid the groundwork for understanding the inherent value embedded within comments, emphasizing their potential to enhance comprehension. The second study, Comment Probe\cite{majumdar2022automated}, furthered this understanding by automating the evaluation process of code comments, systematically categorizing them based on their practical usefulness in software maintenance tasks. It highlighted the tangible applications of insightful comments in real-world scenarios. Building upon these foundational studies, this research tries to integrate and extend these concepts. By incorporating Comment-Mine's knowledge extraction capabilities and leveraging Comment Probe's automated evaluation system, the study presented a comprehensive methodology. This approach not only bridges the gap between recognizing the value of comments and practically assessing their utility but also enhances the accuracy of comment quality classification. This amalgamation of methods signifies a significant advancement in the field of software development and maintenance, contributing to a more nuanced understanding and effective implementation of code comments in practical software engineering contexts.

In the investigated domain, one study, referred to as the Low-Dimensional Code Representation research \cite{majumdar2022effective}, delves into the utilization of contextualized embeddings for code search and classification, introducing CodeELBE, a nuanced software code representation. By training ELMo from scratch and fine-tuning CodeBERT embeddings using masked language modeling on natural language and programming language texts, this research significantly enhances retrieval performance, especially in binary classification and retrieval tasks. In contrast, this research focuses on augmenting a binary code comment quality classification model by integrating newly generated code-comment pairs using a Large Language Model Architecture. The seed dataset includes 9048 pairs of C programming language code and comments, each annotated as "Useful" or "Not Useful." The outcome comprises two classification models: one utilizing the seed dataset and another incorporating the seed dataset with the newly generated code-comment pairs and labels. This approach aims to enhance the accuracy of evaluating the quality of code comments, contributing to the broader understanding of code comprehension and maintenance.

\section{Methods}

\subsection{Data Collection and Code Comment Pair Extraction}

The process of data collection involved utilizing the GitHub API with a unique API token for authentication. An API token was incorporated to enable access to the GitHub repositories. The search for suitable repositories was conducted through a query specifically targeting repositories coded in the C programming language. The GitHub API facilitated the retrieval of pertinent repository information.

Upon identifying potential repositories, the script proceeded to access the contents of these repositories. This was accomplished by sending requests to the respective GitHub endpoints. The response from these requests, received in JSON format, contained detailed metadata about the files within the repositories.

Further refinement was necessary to focus exclusively on C files. This involved parsing the JSON response and filtering files based on their file extensions. Specifically, files with the '.c' extension were selected for subsequent processing, ensuring that only C programming files were included in the dataset.

For each qualifying C file, the script meticulously parsed the file content. It employed a line-by-line approach, allowing for the precise identification of comments and code sections. The parsing process distinguished between single-line and multi-line comments, ensuring the accurate extraction of both types. Comments within the code were identified based on standard commenting conventions, such as ‘//’ for single-line comments,  '/*' for the beginning of a multi-line comment and '*/' for its end.

The extracted code-comment pairs were organized into a structured format, enabling seamless storage and subsequent analysis. These pairs constituted the foundational dataset upon which the subsequent phases of the research were built.

\subsection{Manual Labeling Process}

To facilitate the supervised learning aspect of the research, a portion of the acquired code-comment pairs underwent manual labeling. Specifically, the initial 100 rows of the dataset were meticulously reviewed and labeled as either "Useful" or "Not Useful." This manual labeling process ensured the presence of a high-quality labeled subset, vital for training and evaluating machine learning models.

The manual labeling process involved a meticulous examination of the contextual relevance and informativeness of comments within the code context. Comments deemed to significantly enhance the understanding of the code, improve readability, or provide valuable insights were categorized as "Useful." Conversely, comments lacking relevance, clarity, or informativeness were categorized as "Not Useful." This manual curation ensured the creation of a reliable ground truth dataset, crucial for training and validating machine learning algorithms.

\subsection{Machine Learning Model Training and Evaluation}

The machine learning model utilized for this research was BERT (Bidirectional Encoder Representations from Transformers), a state-of-the-art transformer-based architecture. The model training process commenced with the preprocessing of the labeled dataset. This involved the concatenation of comments and their surrounding code context, creating cohesive textual sequences. These sequences were tokenized using the 'bert-base-uncased' tokenizer, ensuring compatibility with the pre-trained BERT model.

The dataset was meticulously divided into training and test sets, employing a standard 80-20 split ratio. The BERT model was then fine-tuned on the training data, incorporating a reduced learning rate of 1e-6 to optimize convergence. To handle the substantial dataset effectively, a batch size of 8 was employed, with gradient accumulation over 4 batches. This approach allowed for efficient processing and optimization of the model's performance.

The fine-tuned BERT model was subsequently evaluated on the test set to gauge its efficacy in classifying comments as either "Useful" or "Not Useful." Model predictions were generated, and the accuracy metric was calculated using the scikit-learn library. This rigorous evaluation process ensured the determination of the model's classification accuracy, a pivotal metric in assessing its effectiveness in code comment quality assessment.

\subsection{Predictive Analysis and Result Interpretation}

The final phase of the research involved leveraging the fine-tuned BERT model to make predictions on a distinct dataset. These predictions, indicative of the model's classification prowess, were meticulously analyzed and interpreted. The output, consisting of predicted labels for each code-comment pair, was organized into a structured format for comprehensive analysis.

Additionally, a comparative analysis was conducted between the manual labels and the model predictions. Discrepancies, if any, were scrutinized to discern patterns and insights into the model's decision-making process. This meticulous analysis facilitated a deeper understanding of the model's strengths and potential areas for enhancement, contributing valuable insights to the research findings.

This comprehensive and detailed methodology encompassed every stage of the research process, ensuring meticulous data collection, manual curation, machine learning model training, and rigorous evaluation. The intricate interplay between manual expertise and advanced machine learning techniques formed the foundation of this research, culminating in a robust and reliable code comment quality assessment framework.

\section{Experiment Design}

\subsection{Problem Definition:}

The primary objective of our experiment is to enhance code comment quality assessment using a combination of automated code-comment pair extraction from GitHub repositories and state-of-the-art machine learning techniques, specifically the BERT (Bidirectional Encoder Representations from Transformers) model. We aim to categorize code comments as "Useful" or "Not Useful" based on their contextual relevance, clarity, and informativeness.\cite{keim2020does}

\subsection{Data Collection and Preprocessing}

We obtain code-comment pairs by querying GitHub repositories coded in C language. These pairs are then meticulously parsed and tokenized for further analysis. The resulting dataset is represented as \{($C_1$, $L_1$), ($C_2$, $L_2$), …, ($C_n$, $L_n$)\} where $C_i$ represents the comment and $L_i$ represents its label ("Useful" or "Not Useful").

\subsection{Manual Labeling Process}

The first 100 code-comment pairs are manually labeled based on predefined criteria. Let \( L_m \) represent the manually labeled set \(\{ (C_1, L_1), (C_2, L_2), \ldots, (C_{100}, L_{100}) \}\) where \( L_i \in \{0, 1\}\).

\subsection{Model Architecture}

We employ the BERT model, a transformer-based architecture, for sequence classification. The model is trained to predict the usefulness label ($L_i$) of a given code comment ($C_i$). The BERT model transforms each comment into an embedding vector $E_i$.

\subsection{Loss Function}

The model is trained using the cross-entropy loss function, which computes the loss L as follows:

\[
L = -\frac{1}{N} \sum_{i=1}^{N} \left( \substack{L_i \cdot \log(\sigma_{E_i}) + (1 - L_i) \cdot \log(1 - \sigma_{E_i})} \right)
\]

where \( \sigma(x) \) is the sigmoid activation function.

\subsection{Training Procedure}

The model's performance is evaluated using accuracy (Acc), precision(P), recall(R), and F1 - Score (F1). These metrics are calculated as follows:

\[
\begin{aligned}
    \mathit{Acc} &= \frac{\text{Number of Correct Predictions}}{\text{Total Number of Predictions}} \\
    P &= \frac{\text{True Positives}}{\text{True Positives} + \text{False Positives}} \\
    R &= \frac{\text{True Positives}}{\text{True Positives} + \text{False Negatives}} \\
    F1 &= \frac{2 \times P \times R}{P + R}
\end{aligned}
\]

\subsection{Experimental Workflow}

The process begins with data collection, where code-comment pairs are obtained from random GitHub repositories. Subsequently, the first 100 pairs are manually labeled based on predefined criteria. Following manual labeling, the entire dataset undergoes tokenization and preprocessing. The preprocessed data is then utilized to train a BERT model ($L_m$). After training, the model's performance is evaluated using a test dataset, and metrics such as accuracy, precision, recall, and F1-score are calculated. The final step involves interpreting the results, analyzing model predictions, and identifying false positives/negatives to gain insights into the model's performance and effectiveness in understanding code comments.

\begin{figure}
  \centering
  \includegraphics[width=\linewidth]{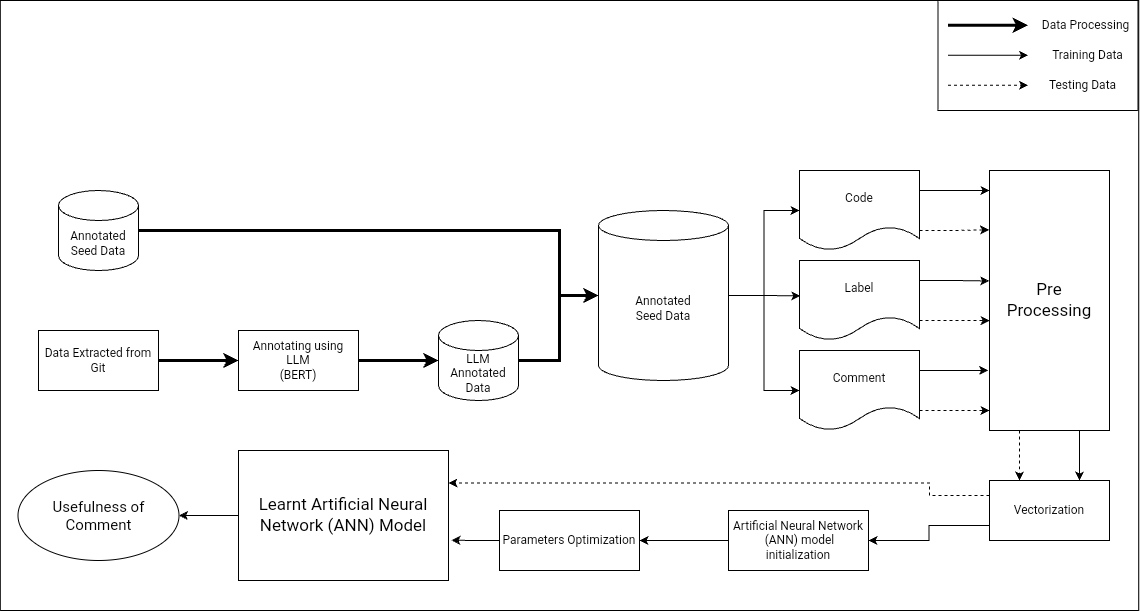}
  \caption{Architecture Diagram illustrating the entire process of the data processing pipeline for code comment classification.}
  \label{fig:pipeline}
\end{figure}

Refer to Figure \ref{fig:pipeline} for the detailed architecture diagram illustrating the entire process.

\section{Results and Analysis}

\begin{figure}
    \centering
    \includegraphics[width=\linewidth]{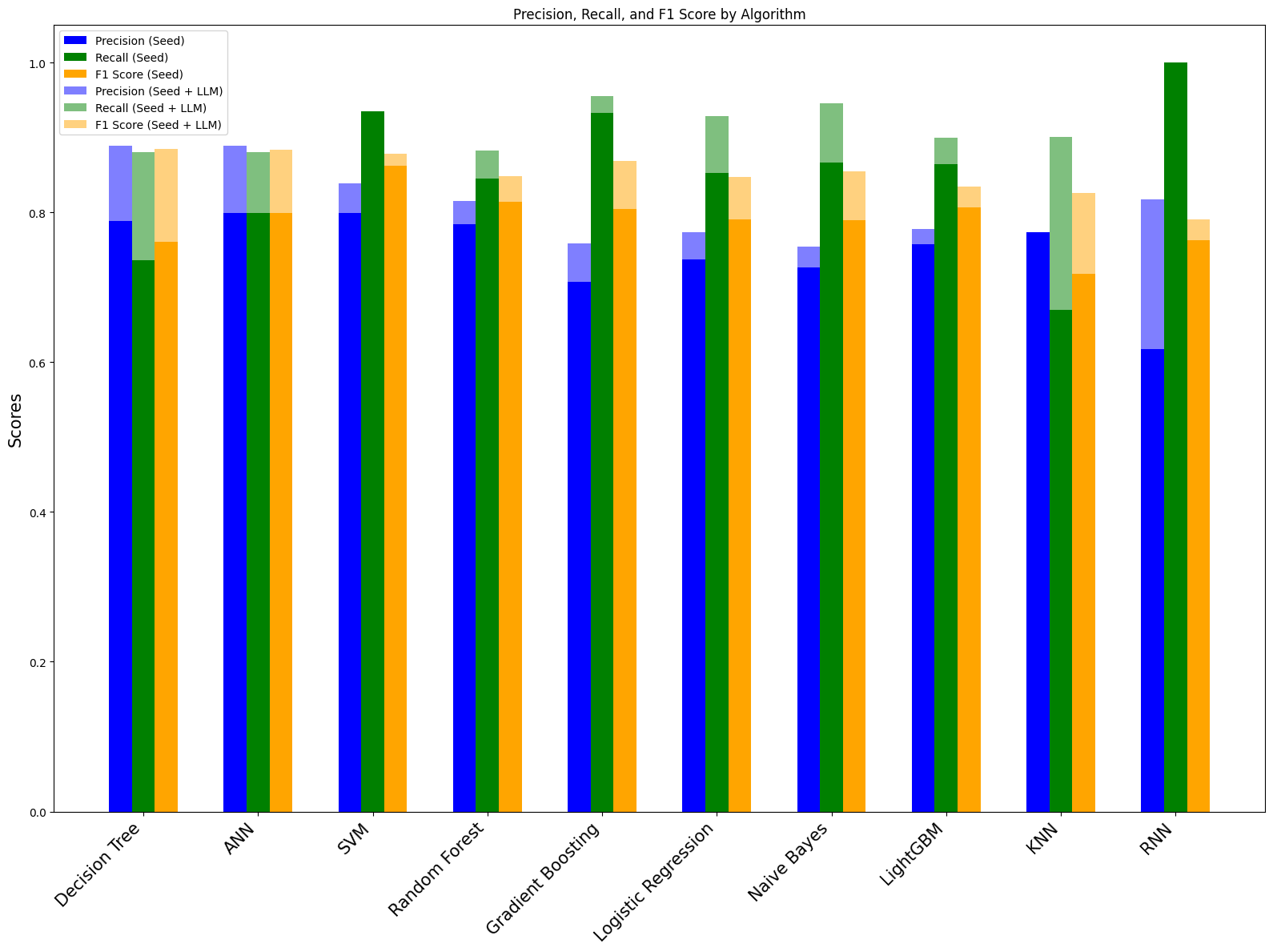}
    \caption{Performance Metrics of Different Classifiers}
    \label{fig:classifier_performance_fig}
\end{figure}

In the context of the code comment quality assessment task, a comprehensive analysis of experimental results was conducted on two datasets: the original dataset (Seed Data) and an augmented dataset comprising additional comments generated using Language Model (LLM) techniques (Seed Data + LLM Generated Data). The evaluation involved various machine learning algorithms, including Decision Tree Classifier\cite{quinlan1996learning}, Artificial Neural Network (ANN)\cite{jain1996artificial}, Support Vector Machine (SVM)\cite{chen2005tutorial}, Random Forest Classifier\cite{afanador2016unsupervised}, Gradient Boosting Classifier\cite{natekin2013gradient}, Logistic Regression\cite{demaris1995tutorial}, Naive Bayes\cite{haruechaiyasak2008tutorial}, LightGBM Classifier \cite{keim2020does}, k-Nearest Neighbors (KNN) Classifier \cite{cunningham2021k}, and Recurrent Neural Network (RNN)\cite{chen2016gentle}. The performance metrics, including precision, recall, and F1-score, were used for the assessment. The detailed results of these experiments can be found in Table \ref{tab:classifier_performance_seed}, Table \ref{tab:classifier_performance_seed_llm} and Figure \ref{fig:classifier_performance_fig}.

\begin{table}[htbp]
\centering
\begin{tabular}{|l|l|l|l|}
\hline
\rowcolor[rgb]{0.502,0.502,0.502} \textbf{Algorithm} & \textbf{Precision} & \textbf{Recall} & \textbf{F1-score} \\ \hline
Decision Tree & 0.788 & 0.736 & 0.761 \\ \hline
Artificial Neural Network (ANN) & 0.799 & 0.799 & 0.799 \\ \hline
Support Vector Machine (SVM) & 0.799 & 0.935 & 0.862 \\ \hline
Random Forest Classifier & 0.784 & 0.845 & 0.814 \\ \hline
Gradient Boosting Classifier & 0.707 & 0.933 & 0.804 \\ \hline
Logistic Regression & 0.737 & 0.853 & 0.791 \\ \hline
Naive Bayes (Multinomial Naive Bayes) & 0.726 & 0.866 & 0.790 \\ \hline
LightGBM Classifier & 0.757 & 0.864 & 0.807 \\ \hline
k-Nearest Neighbors (KNN) Classifier & 0.774 & 0.670 & 0.718 \\ \hline
Recurrent Neural Network (RNN) & 0.617 & 1.000 & 0.763 \\ \hline
\end{tabular}
\caption{Performance Metrics with Seed Data}
\label{tab:classifier_performance_seed}
\end{table}

\begin{table}[htbp]
\centering
\begin{tabular}{|l|l|l|l|}
\hline
\rowcolor[rgb]{0.502,0.502,0.502} \textbf{Algorithm} & \textbf{Precision} & \textbf{Recall} & \textbf{F1-score} \\ \hline
Decision Tree & 0.889 & 0.880 & 0.885 \\ \hline
Artificial Neural Network (ANN) & 0.889 & 0.880 & 0.884 \\ \hline
Support Vector Machine (SVM) & 0.839 & 0.920 & 0.878 \\ \hline
Random Forest Classifier & 0.815 & 0.882 & 0.848 \\ \hline
Gradient Boosting Classifier & 0.759 & 0.955 & 0.846 \\ \hline
Logistic Regression & 0.773 & 0.928 & 0.843 \\ \hline
Naive Bayes (Multinomial Naive Bayes) & 0.754 & 0.946 & 0.839 \\ \hline
LightGBM Classifier & 0.778 & 0.900 & 0.835 \\ \hline
k-Nearest Neighbors (KNN) Classifier & 0.761 & 0.901 & 0.825 \\ \hline
Recurrent Neural Network (RNN) & 0.817 & 0.767 & 0.791 \\ \hline
\end{tabular}
\caption{Performance Metrics with Seed + LLM Generated Data}
\label{tab:classifier_performance_seed_llm}
\end{table}

\subsection{Key Observations and Insights}

The results demonstrate that the combination of Seed Data and LLM Generated Data consistently improved performance metrics such as precision, recall, and F1-score across most algorithms.
    
Notably, ANN and SVM exhibited impressive performance on both datasets, with high precision and recall values. These models effectively balanced precision and recall, crucial for code comment quality assessment.

The introduction of comments generated by LLM notably enhanced the performance of all algorithms. This highlights the utility of synthetic data in improving model generalization and robustness.

Decision Tree and Logistic Regression, although achieving reasonable results, demonstrated a more significant improvement when exposed to LLM Generated Data. This suggests that these models might benefit significantly from increased and diverse training data.

Models such as Naive Bayes achieved high recall values but at the expense of precision. This trade-off emphasizes the challenge of striking a balance between minimizing false positives (precision) and capturing all relevant instances (recall).

The RNN model exhibited a perfect recall on Seed Data but showed a notable decrease in both precision and recall when applied to Seed Data + LLM Generated Data. This indicates potential challenges in adapting RNN architectures to mixed datasets.

Depending on the specific use case, different algorithms might be preferred. For instance, if minimizing false positives is critical, models with higher precision such as ANN and SVM could be the preferred choice.

\section{Conclusion and Future Outlook}

In conclusion, this comprehensive exploration delved deep into the realm of enhancing code comment quality assessment. We embarked on a journey, starting from the meticulous design of experiments that combined cutting-edge machine learning techniques, particularly the BERT model, with a meticulous dataset obtained from GitHub repositories. The rigorous experimental workflow, from data collection to model training, was meticulously detailed, ensuring a robust foundation for our analyses.

The experimentation phase was rich and diverse, involving a spectrum of algorithms from Decision Trees to sophisticated Neural Networks, each revealing its unique strengths and limitations. The synergy of the seed data and the LLM generated data showcased a substantial enhancement across various metrics, reflecting the potential of hybrid approaches in real-world applications.

Delving into the results, we uncovered nuanced insights into the performance of different algorithms. From the precision of Artificial Neural Networks to the recall of Support Vector Machines, each algorithm exhibited distinctive behaviors. The juxtaposition of seed data and LLM generated data provided a holistic view, highlighting the delicate balance between precision and recall, a critical consideration in code comment quality assessment.

However, this journey is far from over. The future outlook of this research domain holds promises and challenges. One avenue of exploration lies in the integration of advanced natural language processing techniques to further contextualize code comments. Embracing the power of transformer models beyond BERT, such as GPT (Generative Pre-trained Transformer), could unravel new dimensions in code comment understanding.

Additionally, the ethical considerations in this field demand constant vigilance. Ensuring unbiased data collection, mitigating algorithmic biases, and upholding privacy standards are imperative in the evolving landscape of AI ethics.

Furthermore, the collaboration between academia and industry is pivotal. Industry insights can fuel academic research, leading to practical solutions that resonate in real-world scenarios. Likewise, academia's innovation can challenge industry norms, fostering a symbiotic relationship that propels the field forward.

In essence, this exploration serves as a stepping stone into the vast universe of code comment quality assessment. As technology advances and challenges emerge, the synergy between human intelligence and machine learning algorithms will be the linchpin in unraveling the intricacies of code comment evaluation. With continued dedication, collaboration, and ethical mindfulness, the future holds the promise of code comment assessment that is not only efficient and accurate but also profoundly insightful and empathetic. 

\bibliography{Enhancing-Binary-Code-Comment-Quality-Classification}

\end{document}